\documentclass[preprint]{aastex6}

\slugcomment{Accepted for publication in the Astrophysical Journal}

\begin{document}

\title{Internal Structures of Molecular Clouds in the LMC
Revealed by ALMA}

\author{Tsuyoshi Sawada\altaffilmark{1,2}}
\affil{Joint ALMA Observatory,
  Alonso de C\'{o}rdova 3107, Vitacura,
  Santiago 763-0355, Chile}

\author{Jin Koda}
\affil{Department of Physics and Astronomy,
  Stony Brook University, Stony Brook, NY 11794-3800, USA}

\and

\author{Tetsuo Hasegawa}
\affil{National Astronomical Observatory of Japan,
  National Institutes of Natural Sciences,
  2-21-1 Osawa, Mitaka, Tokyo 181-8588, Japan}

\altaffiltext{1}{National Astronomical Observatory of Japan,
  National Institutes of Natural Sciences}
\altaffiltext{2}{sawada.tsuyoshi@nao.ac.jp}

\begin{abstract}

We observed five giant molecular clouds (GMCs) in the Large Magellanic
Cloud (LMC) in the $^{12}{\rm CO}$ $J=1\mbox{--}0$ line using the
Atacama Large Millimeter/submillimeter Array (ALMA).
The sample includes four GMCs with some signs of star formation --
either YSOs, \ion{H}{2} regions, and/or young clusters --
and one quiescent GMC without any sign of massive star formation.
The data from the ALMA 12 m, 7  m, and Total-Power arrays are
jointly deconvolved to obtain high-fidelity images
at high spatial resolution ($3\arcsec = 0.7\;{\rm pc}$).
The four star-forming GMCs show very complex structures with clumps
and filaments.  The quiescent GMC shows a relatively diffuse,
extended emission distribution without prominent clumps or filaments.
This difference is similar to that between structured molecular gas
in Milky Way spiral arms and unstructured gas in the inter-arm regions.
We characterize the difference with the brightness
distribution function and brightness distribution index.
In conjunction with other ALMA studies of GMCs in the LMC,
the five GMCs tentatively form an evolutionary trend:
from less structured, quiescent GMCs to more structured,
actively star-forming GMCs.
A future ALMA study will be able to map molecular clouds over
the LMC and reveal the evolutionary sequence of molecular clouds.
\end{abstract}


\section{Introduction} \label{sec:intro}

Molecular clouds are the site of star formation.
Their internal structures and evolution are 
the key to understanding the process of star formation.
Our recent studies of molecular gas in the Milky Way (MW) revealed
the structural evolution of the molecular gas
when a 1 pc scale is resolved.
\citet{sawada2012, sawada2012let} found that
the gas in the MW spiral arms is {\it structured}
(i.e., bright and spatially confined, parsec-sized, emission is prominent),
while {\it unstructured} gas (i.e., faint and diffuse) dominates
in the inter-arm regions.
A similar trend has been found among local star-forming and
quiescent molecular clouds \citep[e.g., ][]{kainulainen2009}.
In order to characterize these internal structures,
\citet{sawada2012} introduced two simple tools:
the brightness distribution function (BDF) and
brightness distribution index (BDI; see Section \ref{sec:BDF}).

An extension of such study to external galaxies was,
however, limited by the low spatial resolution, typically comparable
to the typical size of giant molecular clouds (GMCs),
several tens of parsecs, or even worse
\citep[e.g., ][]{koda2009, schinnerer2014}.
Recently, the Atacama Large Millimeter/submillimeter Array
\citep[ALMA;][]{hills2010} has revolutionized
such studies in the Large and Small Magellanic Clouds
\citep[LMC and SMC; e.g., ][]{indebetouw2013}.
It can resolve the spatial scale relevant to star formation
\citep[$\sim 1$ pc; e.g.,][]{lada2003} at the distances
of the LMC and SMC.
Most studies so far have focused on the areas around star-forming
regions, e.g., 30 Doradus, the most active starburst region in the
Local Group \citep{indebetouw2013, fukui2015, saigo2017, nayak2018},
and the \ion{H}{2} region N55, which is located within the largest
supergiant shell (SGS) in the LMC \citep{naslim2018}.
Molecular clouds in these star-forming environments
show complex internal structures,
such as clumps and filaments.
\citet{muraoka2017} also showed the presence of small
clumps in the star-forming cloud N83C in the SMC.
On the contrary,
\citet{wong2017} found a less structured emission distribution
in a quiescent molecular cloud in the LMC.

This paper aims to bridge the gap between these two extremes,
clouds with very active star formation and those in the quiescent phase,
and investigate the evolutionary trend.
We observed five clouds in the LMC, which show a range of star formation
activity.
The properties of the interstellar medium and star formation in the LMC
have been thoroughly studied at various wavelengths, thanks to its
proximity \citep[$\approx 50$ kpc;][]{pietrzynski2013} and nearly
face-on geometry \citep[$i\approx 35\arcdeg$;][]{vandermarel2001}.
For this study, it was essential to recover both compact
and extended structures.
Thus, we combine and jointly deconvolve the data from the ALMA 12 m,
7 m, and Total-Power (TP) arrays.

The presence of CO-dark ${\rm H_2}$ is being actively
discussed \citep{greiner2005,planck2011}.
It should be predominantly in the outskirts of GMCs, where the CO is
photodissociated by ultraviolet radiation from the outside, while
${\rm H_2}$ is optically thick against Lyman-Werner photons and
protected.
The ALMA observations and discussions in this paper are based on CO
observations, and hence are about CO-bright parts of GMCs,
not the outskirts.

\section{Target Selections} \label{sec:targets}

The goal of this paper is to study the evolutionary sequence in
cloud structure from relatively quiescent to
actively star-forming molecular clouds.
Obviously, we can only get a peek at the trend with the limited
amount of observing time.
Our sample of clouds and field coverages for each cloud are therefore
not complete by any means.
Nevertheless, we selected five molecular clouds based on
the cloud evolutionary stages defined by \citet{kawamura2009}.
Their classification is based on the master catalog by
\citet{fukui2008}, which consists of 272 clouds in the LMC from
the $^{12}{\rm CO}$ $J=1\mbox{--}0$ survey with the NANTEN 4~m
telescope at a $2\farcm 6$ ($\approx 40$ pc) resolution.
They classified the GMCs into three phases of evolution:
Types I (no sign of massive star formation),
II (associated with \ion{H}{2} regions, i.e., at an early stage
of star formation), and III (associated with \ion{H}{2} regions
and stellar clusters, i.e., at a late stage of star formation).
We selected five GMCs at three evolutionary stages; i.e.,
GMCs 2, 55, and 225 from Type I, GMC 216 from Type II, and
GMC 197 from Type III (see Table \ref{tbl:params}).

\begin{figure*}
\plotone{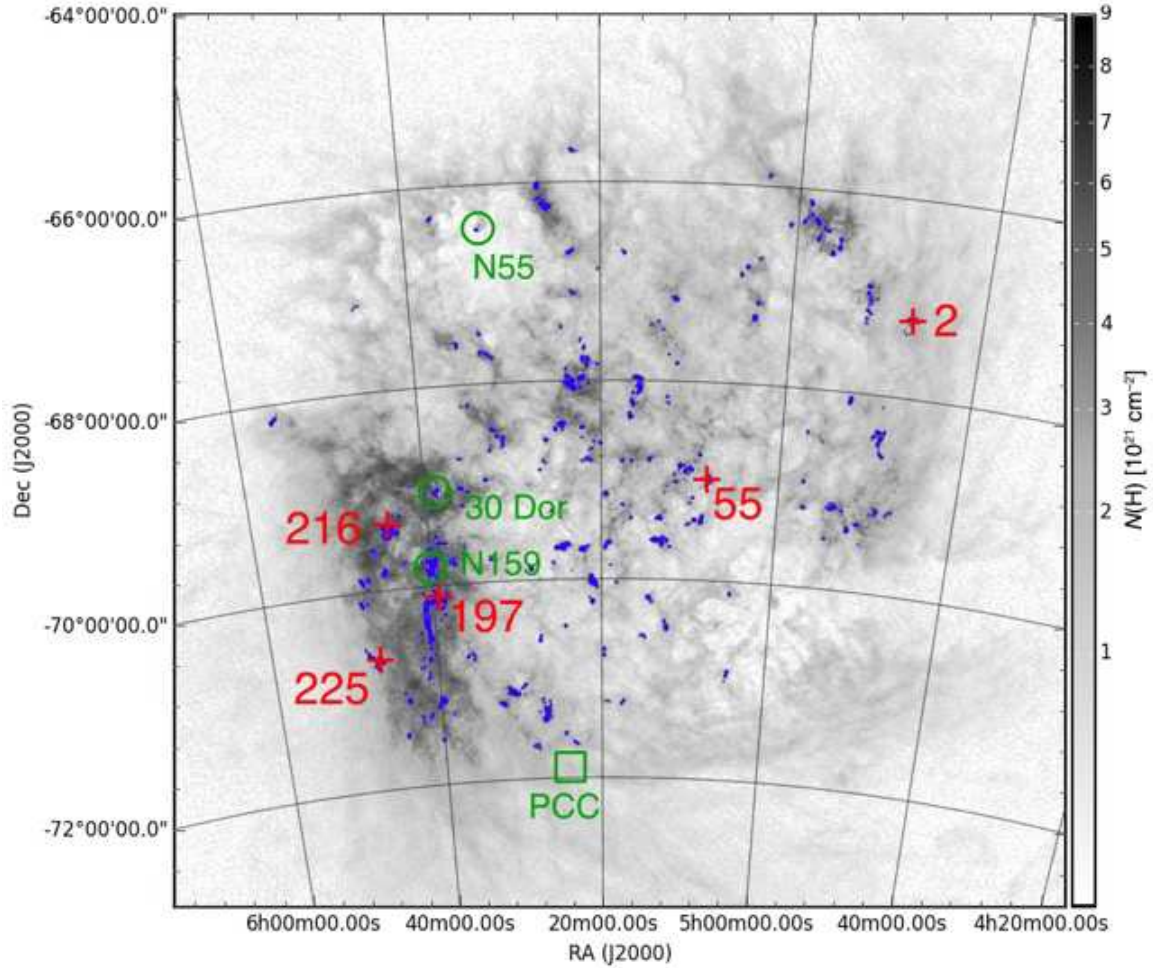}
\caption{The positions of the target GMCs are shown as red crosses
overlaid on the \ion{H}{1} \citep[grayscale;][]{kim2003,staveley-smith2003}
and CO \citep[blue;][]{wong2011} distributions.
Green circles and squares are the locations of the GMCs with
ALMA observations published so far (see the text).}
\label{fig:overall}
\end{figure*}

Figure \ref{fig:overall} illustrates the distribution of our GMCs, 
as well as the other clouds in the literature.
The green circles show the locations of \ion{H}{2} regions
\citep{henize1956} whose surrounding molecular gas was observed with
ALMA: 30 Doradus, N159, and N55.
These clouds show clumps and filaments
\citep{indebetouw2013, fukui2015, saigo2017, naslim2018, nayak2018}.
The quiescent cloud, the ``Planck cold cloud'' (PCC) located
near the southern edge of the LMC (green square)
was also observed with ALMA \citep{wong2017}.
Our five GMCs, 2, 55, 225, 216, and 197,
are distributed from the edge of the disk to near 30 Doradus.
We note that GMCs are sometimes referred to by the names of
nearby \ion{H}{2} regions (e.g., N55, N159), while our GMC IDs
(GMC 2, 55, 225, 216, and 197) are from \citet{fukui2008}.

\subsection{Considerations}

Our target selection and observation strategies were built solely
on \citet{kawamura2009} and \citet{fukui2008}.
Their results were based on the best information available at the
time of their work.
Of course, our knowledge has improved, and one may be tempted to
modify their GMC classification with respect to star formation.
Table \ref{tbl:params} lists some other parameters that characterize
the star formation activity of our GMCs, in addition to their
Kawamura's types based on their association with \ion{H}{2} regions
and young stellar clusters.
For example, the number of young stellar objects (YSOs) within the
30 and 45 pc radii from the center of our ALMA mosaics are zero for
GMC 2, but one to four for the other GMCs.
With this criterion, GMC 2 (Type I) is
the only GMC in our sample that has no sign of star formation,
although GMCs 55 and 225 are also classified as Type I.
For this, we counted only the YSOs from \citet{seale2014}
with a high likelihood (``probable'' ones) of being in the LMC.

The average dust temperature within a field of view (FoV) traces
the radiation field. It is also higher ($T_{\rm dust}>20$ K) for
GMCs 55, 197, and 216 than for GMCs 2 and 225 \citep{gordon2014}.
The 8 $\micron$ flux is mainly from the interstellar polycyclic
aromatic hydrocarbons (PAHs) excited by the background radiation.
Hence, it also indicates the strength of the background radiation field.
The background 8 $\micron$ flux in the table is calculated from
the {\it Spitzer Space Telescope} 8 $\micron$ images \citep{meixner2006}
after removing point sources (i.e., stars).
Again, it is higher for GMCs 55, 197, and 216 than for GMCs 2 and 225
(Figure \ref{fig:intandmax1}).

Depending on the adopted indicator of star formation activity,
the classification of the GMCs may be altered. 
This study has this type of uncertainty.
From the above parameters, GMC 2 is the only target that does not
show any sign of star formation. The other four GMCs show
some sign, though some may be at an earlier stage of star formation
than the others.

In addition, our ALMA FoVs were set around the peak positions
in the CO $J=1\mbox{--}0$ maps at the $2\farcm 6$ resolution
\citep{kawamura2009}.
The follow-up CO $J=1\mbox{--}0$ observations at a higher
resolution \citep[$45\arcsec$; ][]{wong2011} resolved structures
further. It turned out that in some cases, the emission peaks at 
the $2\farcm 6$ resolution are not the peaks at the $45\arcsec$
resolution (Figure \ref{fig:intandmax1}).
Therefore, our ALMA FoVs are not optimal in terms of the peak
positions in the $45\arcsec$ resolution maps.
Despite these retrospective considerations, the new ALMA data
presented here show a tentative trend of cloud evolution.

\begin{deluxetable*}{ccccccccc}
\tablecolumns{7}
\tablewidth{0pc}
\tablecaption{Parameters of the Sample GMCs\label{tbl:params}}
\tablehead{
  \colhead{GMC ID\tablenotemark{a}} &
	\colhead{$R$\tablenotemark{a}} &
    \colhead{$M_{\rm CO}({\rm all})$\tablenotemark{a}} &
    \colhead{Type\tablenotemark{b}} &
    \colhead{$N_{\rm YSO,30/45}$\tablenotemark{c}} &
    \colhead{$T_{\rm dust}$\tablenotemark{d}} &
    \colhead{$S_{8\;{\rm \mu m}}$\tablenotemark{e}} &
    \colhead{$f_{\rm mol}$\tablenotemark{f}} &
    \colhead{SGS Assoc.\tablenotemark{g}}\\
  \colhead{} &
    \colhead{(pc)} &
    \colhead{($10^5M_\sun$)} &
    \colhead{} &
    \colhead{} &
    \colhead{(K)} &
    \colhead{(Jy)} &
    \colhead{} &
    \colhead{}
}
\startdata
2   &  97 &  10 & I   & 0/0 & 16.5 & 0.59 & 0.79 & No \\
55  &  39 &   5 & I   & 1/3 & 20.4 & 1.07 & 0.83 & Rim \\
197 & 220 & 100 & III & 1/2 & 22.4 & 1.80 & 0.86 & Rim \\
216 &  80 &  20 & II  & 2/4 & 24.4 & 1.51 & 0.89 & Inside \\
225 &  73 &  10 & I   & 1/1 & 17.8 & 0.74 & 0.70 & No \\
\enddata
\tablenotetext{a}{From the NANTEN survey \citep{fukui2008}}
\tablenotetext{b}{Type I -- no sign of massive star formation;
Type II  -- associated with \ion{H}{2} regions;
and Type III  -- associated with \ion{H}{2} regions and stellar clusters;
\citet{kawamura2009}}
\tablenotetext{c}{The number of ``probable'' YSOs \citep{seale2014}
within 30 and 45 pc radii from the center of our FoVs.}
\tablenotetext{d}{\citet{gordon2014}}
\tablenotetext{e}{\citet{meixner2006}}
\tablenotetext{f}{The molecular gas fraction
\citep[\ion{H}{1} from][; CO from this work]{kim2003,staveley-smith2003}.
A CO-to-${\rm H_2}$ conversion factor of
$7\times 10^{20}\;{\rm cm^{-2}\,(K\,km\,s^{-1})^{-1}}$
is assumed \citep{fukui2008}}
\tablenotetext{g}{The SGS identification by \citet{dawson2013}
is adopted}
\end{deluxetable*}

\section{Observations} \label{sec:obs}

The target five GMCs were observed with ALMA using the Band 3 receivers
\citep{claude2008} as the Early Science Cycle 1 project 2012.1.00641.S.
The observations using the 12 m array, which consisted of twenty-seven 12 m
antennas and the 64-input correlator \citep{escoffier2007},
were made in 2013 December.
A 469 MHz wide, 3840-channel, dual-polarization spectral window
($244\;{\rm kHz}= 0.64\;{\rm km\,s^{-1}}$ resolution) was placed
at the frequency of the CO $J=1\mbox{--}0$ line (rest frequency
115.271 GHz) in one of the four basebands.
Three 2 GHz wide low spectral resolution spectral windows were set up
in the remaining basebands as a serendipitous search for continuum
emission, but resulted in no detections.
The FoV of $\approx 2\farcm 5 \times 2\farcm 5$ for each GMC
was covered by a 27-pointing mosaic.
The total on-source integration time per object was typically 540 s
(20 s per mosaic pointing),
and the $u\mbox{--}v$ distance typically ranged from 14 to 450 m.
The gain calibration was made by observing the QSO J0635$-$7516 at
a typical interval of 8 minutes.
Either Uranus or the radio galaxy J0519$-$4546, whose flux was
monitored by the observatory, was observed in each execution
as the flux calibrator.
The typical system noise temperature $T_{\rm sys}$ at the frequency
of the CO line was 170 K.

The shorter $u$--$v$ distance range was fulfilled by observations
using the 7 m array, a part of the Atacama Compact Array
\citep[ACA;][]{iguchi2009}.
These observations were done in 2013 November--December and 2014 April
using eight to eleven 7 m antennas and the ACA correlator \citep{kamazaki2012}.
The spectral setup was equivalent to that for the 12 m array.
The number of mosaic pointings was 10 for each GMC.
The typical total on-source integration time and $u$--$v$ distance were
1000 s (100 s per mosaic pointing) and 7--35 m, respectively.
A gain calibrator, QSO J0635$-$7516 in most cases, was observed at
a typical interval of 10 minutes.
Either a solar system object (Mars, Uranus, Callisto, Ganymede, or
Pallas) or J0519$-$4546 was used as the flux calibrator.
The typical $T_{\rm sys}$ was 120 K.

The even shorter $u\mbox{--}v$ spacing was complemented by
filled-aperture (single-dish) observations
with the TP array, which was also a part of the ACA.
The TP array observations were carried out in 2015 May and June
using two or three 12 m antennas and the ACA correlator.
A $225\arcsec \times 225\arcsec$ FoV for each GMC was mapped using
the on-the-fly observing technique.
A line-free reference position was visited before every raster row.
The QSO 3C 279 was also observed to determine the antenna gains
(${\rm Jy\;K^{-1}}$).
The typical total on-source time (per antenna) and $T_{\rm sys}$ were
60 minutes and 140 K, respectively.

\section{Data Reduction} \label{sec:reduction}

Interferometer visibilities and single-dish maps from the ALMA 12 m,
7 m, and TP arrays were delivered by the observatory and were
calibrated using the Common Astronomy Software Applications
\citep[CASA;][]{mcmullin2007}.
We subtracted spectral baselines from the delivered TP map by fitting
straight lines to emission-free channels.
We also applied the correction for the weights of 12 and 7 m
visibilities with the {\tt statwt} task in CASA as described in the
CASA guide\footnote{https://casaguides.nrao.edu/index.php/DataWeightsAndCombination}.

A better-filled $u$--$v$ coverage is always advantageous in deconvolution.
We combined the 12 m, 7 m, and TP data in $u$--$v$ space and
deconvolved them jointly.
We converted the TP map into visibilities by using the Total Power to
Visibilities (\textsc{TP2VIS}) package that runs on the CASA
platform\footnote{https://github.com/tp2vis/distribute/blob/master/README.md}.
This procedure is discussed and tested in depth in our previous paper
\citep{koda2011}.
Here, we briefly describe the essence of this procedure.
\textsc{TP2VIS} generates a Gaussian visibility distribution, so that
when it is Fourier transformed its dirty image and beam represent
the TP map and primary beam, respectively.
The optimal weight of the TP visibilities with respect to 12 and 7 m
ones can be debated.
We adopted the TP weight that corresponds to the rms
noise in the TP map.
This way the weights represent the quality of data properly.

For deconvolution, we used the Multichannel Image Reconstruction,
Image Analysis, and Display (Miriad) software package \citep{sault1995},
instead of CASA.
The {\tt clean} and {\tt tclean} tasks in CASA (as of version 5.3)
use only one dirty beam for a whole mosaic in their minor cycles,
while it varies across the mosaic.
This often causes a divergence in flux and/or artificial stripping
patterns\footnote{https://casaguides.nrao.edu/index.php/M100\_Band3\_Combine\_4.3}.
Therefore, we could not use CASA for imaging.
All of the visibilities were converted to the MIRIAD data format
and were inverted to dirty map and beam with the natural weighting.
We ran the {\tt mossdi2} task for image-based clean.
We set the number of iterations to 4 million and the {\tt gain}
parameter to 0.05.

It is often a problem that the flux of the cleaned map is not
consistent with that of the TP map.
This is because the two components in the cleaned map use two different
beams -- the model component uses a convolution beam, while the
residual component uses the dirty beam.
The areas of the two beams are often not the same, which leads to
errors in brightness and flux.
To circumvent this problem, we used the residual scaling scheme of
\citet{Jorsater1995}.

The convolution beams for the five GMCs are typically an ellipse with
a major axis size of $2\farcs 2\mbox{--}2\farcs 6$.
For analyses on an equal basis, we smoothed the maps to
a common $3\farcs 0$ resolution (0.73 pc) along both the major and
minor axes of the beams.
We use a velocity channel spacing of $1\;{\rm km\,s^{-1}}$,
at which the sensitivity ranges between 0.47 and 0.67 K
(45--66 ${\rm mJy\,beam^{-1}}$).

\section{Results} \label{sec:results}

The CO $J=1\mbox{--}0$ integrated and peak intensity maps are
shown in Figure \ref{fig:intandmax1}.
Each CO $J=1\mbox{--}0$ map covers a $2\farcm 6 \times 2\farcm 8$
($38 \times 41\rm\, pc^2$) region.
A scale bar is in one of the bottom panels.
We also show a pseudo-color image from {\it Spitzer}
(24, 8.0, 3.6$\mu$m in R, G, B, respectively) and contours
from the Mopra CO $J=1\mbox{--}0$ integrated intensity \citep{wong2011}.
YSO candidates are also marked.
The thick white lines indicate the areas of the ALMA CO $J=1\mbox{--}0$
observations.
The centers of our ALMA mosaics were set at the peak positions
in \citet{kawamura2009}, which later turned out to not coincide with
the peaks at the higher-resolution Mopra CO $J=1\mbox{--}0$ map.
The velocity channel maps are presented in Figures
\ref{fig:chmap002}--\ref{fig:chmap225}.

The parameters of the GMCs derived
in this paper are listed in Table \ref{tbl:results}.
We adopt a CO-to-${\rm H_2}$ conversion factor of
$7\times 10^{20}\;{\rm cm^{-2}\,(K\,km\,s^{-1})^{-1}}$
\citep{fukui2008}.
The masses within the ALMA fields range
between 1.1 and $4.9 \times 10^5 M_\sun$;
5--30\% of the total GMC masses in the NANTEN catalog.
Below, we note characteristics of individual GMCs.

\begin{description}

\item[GMC 2]: This cloud is classified as Type I, without sign of
massive star formation, by \citet{kawamura2009}.
Indeed, it does not have YSOs \citep{seale2014}.
From the channel maps,
the total line width (full-width zero intensity; FWZI) is
$\approx 15\;{\rm km\,s^{-1}}$. We find no bright emission ($>$ 10 K).
At the lower velocity (248--254 ${\rm km\,s^{-1}}$) fluffy emission
is distributed over the FoV.
The typical brightness temperature is 3--5 K.
At 255--259 ${\rm km\,s^{-1}}$ a $\approx$ 15 pc blob dominates.
It is rather featureless, i.e., its brightness temperature is
more or less uniform (3--5 K).

\item[GMC 55]: Type I, but having YSOs, and hence star forming.
The total line width (FWZI) is $\approx 25\;{\rm km\,s^{-1}}$.
Possibly, there are two velocity components.
The most prominent structure in the lower velocity range
(245--257 ${\rm km\,s^{-1}}$) is a filament, or
a chain of compact clumps separated by $\approx$ 5 pc,
stretching from the NE to W of the FoV.
The width of the filament, or the typical size of the clumps,
is $\approx$ 2 pc, and the length (in the FoV) is $\approx$ 30 pc.
There is a velocity gradient along the filament
(several ${\rm km\,s^{-1}}$ over 30 pc).
The line width is a few ${\rm km\,s^{-1}}$ at each portion
along the filament.
There might be another filament, or a chain of clumps, near the SW
corner running from the middle of the western edge toward the SE.
At a higher velocity (255--269 ${\rm km\,s^{-1}}$) there is an
elongated ($\approx 10 \times 2$ pc) bright ($\approx 14$ K) clump with
a sharp boundary in the SE of the FoV.
Its line width (FWZI) is $\approx 15\;{\rm km\,s^{-1}}$.

\item[GMC 197]: Type III with both \ion{H}{2} regions and young star
clusters.
The total line width (FWZI) is $\approx 25\;{\rm km\,s^{-1}}$.
There may be two velocity components.
Emission at the bottom of the FoV shows up at 218 ${\rm km\,s^{-1}}$
and this structure persists until 232 ${\rm km\,s^{-1}}$ or so
(15 ${\rm km\,s^{-1}}$ width).
At 224 ${\rm km\,s^{-1}}$, filamentary structures start to appear,
generally elongated in the N--S direction.
There is a bright spot in one of them
(${\rm R.A.} \approx 5^{\rm h} 39^{\rm m} 52^{\rm s}$,
${\rm decl.} \approx -70\arcdeg 06\arcmin 45\arcsec$)
that has $\approx 15\;{\rm km\,s^{-1}}$ line width and a NW-to-SE
velocity gradient.
It looks like filaments are spreading radially from around that spot
at 228--229 ${\rm km\,s^{-1}}$.
The filaments merge into a $\approx$ 20 pc blob seen at 224--234
${\rm km\,s^{-1}}$.
After that blob is gone, another blob appear in the south, with a
several ${\rm km\,s^{-1}}$ line width.

\item[GMC 216]: Type II with \ion{H}{2} regions and YSOs.
The total line width (FWZI) is $\approx 15\;{\rm km\,s^{-1}}$.
At lower velocities (221--225 ${\rm km\,s^{-1}}$),
at least four filamentary structures are apparent and form a web.
Along the eastern boundary of the map, there is one filament stretching
straight from N to S
(${\rm R.A.}\approx 5^{\rm h} 44^{\rm m} 47^{\rm s}$).
Two filamentary structures run from NEE to W and SE to W, and
apparently merge near the western boundary to form a $\approx$ 15 pc
blob with a line width of several ${\rm km\,s^{-1}}$.
These structures evolve into more spatially extended blobs
at higher velocities (226--230 ${\rm km\,s^{-1}}$).
Hence, it is not clear if they are filaments, or parts of blobs
with large velocity widths. They may be shearing surfaces of, e.g.,
cloud--cloud collisions.
There is another shorter, 10 pc scale, filament that runs
in parallel to and north of the NEE--W filamentary structure.
At 225--231 ${\rm km\,s^{-1}}$ the emission mostly fills
the FoV (typically 5--10 K, with some bright spots).

\item[GMC 225]: Type I, but having YSOs and star formation.
The total line width (FWZI) is
$\approx 10\;{\rm km\,s^{-1}}$.
There are two components, one in the N and the other in the S.
One is a $\approx$ 15 pc blob
in the NW (which may extend beyond the FoV).
The velocity range is 213--223 ${\rm km\,s^{-1}}$.
The brightness is moderately high ($\approx 10$ K)
in its northern half and lower ($\approx 5$ K) in the south.
The southern component is at 216--221 ${\rm km\,s^{-1}}$.
It may consist of two filaments running parallel to each other.
They have a roughly 15$\times$2 pc geometry within the FoV.

\end{description}

\begin{figure*}
\plotone{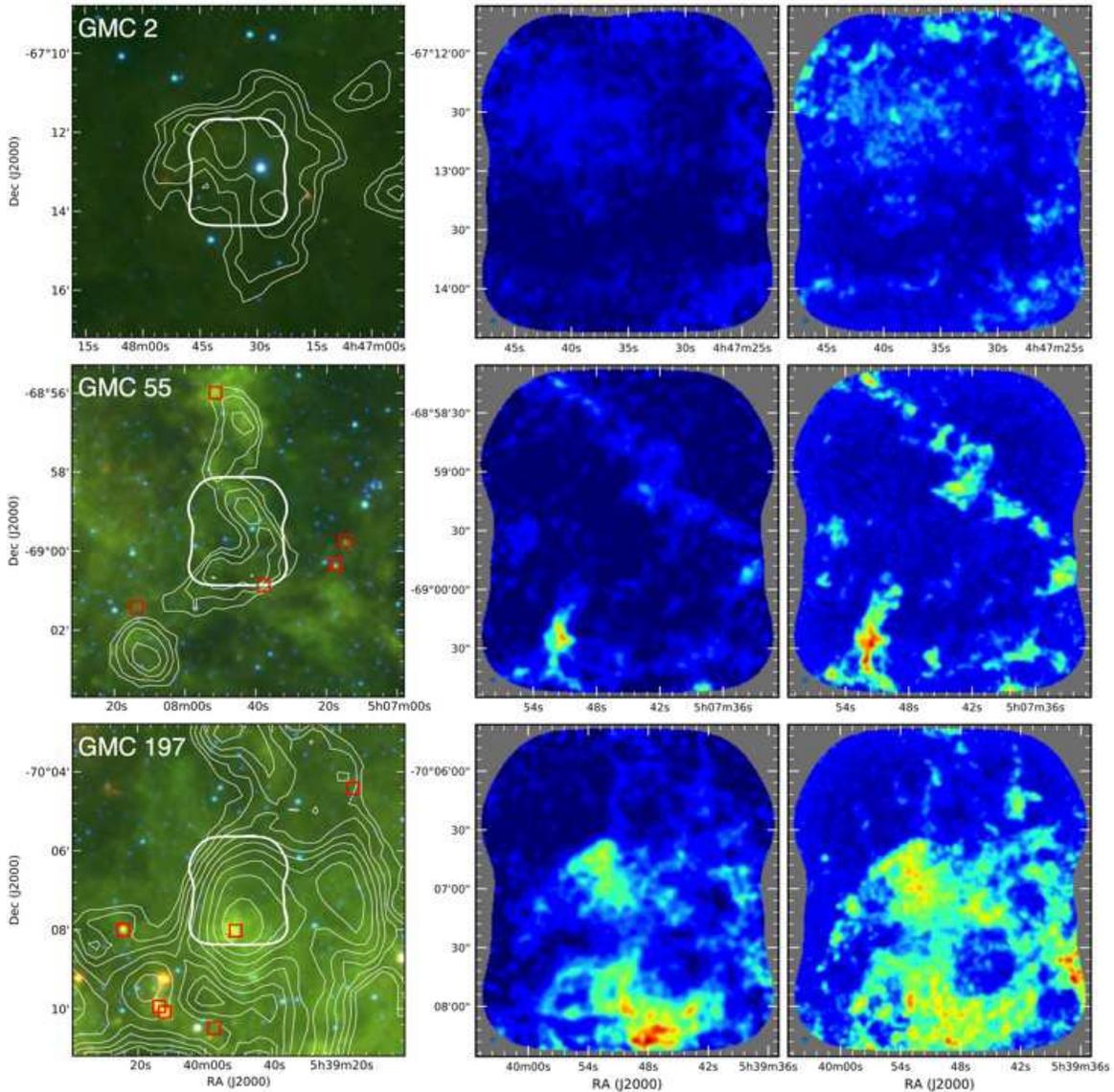}
\caption{Left: Contours of Mopra CO $J=1\mbox{--}0$ integrated intensity
\citep[2, $2\sqrt{2}$, 4, ..., $32\sqrt{2}\;{\rm K\,km\,s^{-1}}$;][]{wong2011}
toward GMCs 2, 55, 197, 216, and 225 (each row),
overlaid on {\it Spitzer} 24--8.0--3.6 $\micron$ three-color images
\citep{meixner2006}.
Thick white lines show the ALMA FoVs.
The red squares indicate the locations of ``probable'' YSOs
\citep{seale2014}.
Middle: The ALMA CO $J=1\mbox{--}0$ integrated intensity maps.
Right: The ALMA CO $J=1\mbox{--}0$ peak intensity maps.
The ALMA FoVs are $\approx 2\farcm 6 \times 2\farcm 8$
(or $38\times 41\;{\rm pc^2}$).
The spatial resolution ($3\arcsec$) is indicated by the blue dots at
the bottom left corner of each panel.}
\label{fig:intandmax1}
\end{figure*}

\begin{figure*}
\addtocounter{figure}{-1}
\plotone{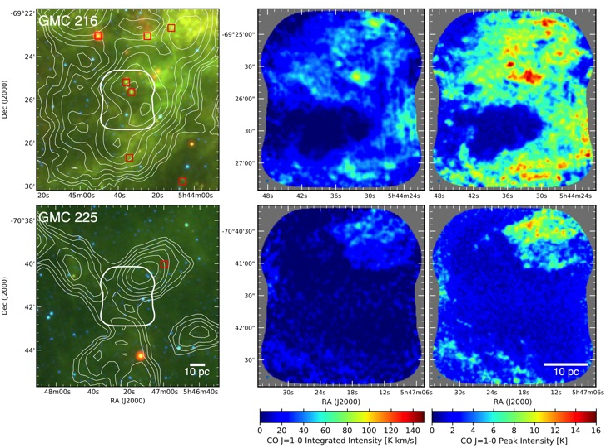}
\caption{{\it Continued}}
\label{fig:intandmax2}
\end{figure*}

\begin{deluxetable*}{cccccccc}
\tablecolumns{7}
\tablewidth{0pc}
\tablecaption{The Observed Quantities for the Sample GMCs
  \label{tbl:results}}
\tablehead{
  \colhead{GMC ID} &
    \multicolumn{2}{c}{Center of FoV} &
    \colhead{RMS\tablenotemark{a}} &
    \colhead{$M_{\rm CO}({\rm ALMA})$\tablenotemark{b}} &
    \colhead{$f_{\rm CO}$\tablenotemark{c}} &
    \colhead{BDI}\\
  \colhead{} &
    \colhead{R.A.(J2000)} &
    \colhead{Decl.(J2000)} &
    \colhead{(K)} &
    \colhead{($10^5M_\sun$)} &
    \colhead{} &
    \colhead{} &
    \colhead{}
}
\startdata
2   & $4^{\rm h} 47^{\rm m} 35^{\rm s}$ &
  $-67\arcdeg 13\arcmin 00\arcsec$ & 0.46 & 1.3 & 0.1 & $-\infty$ \\
55  & $5^{\rm h} 07^{\rm m} 45^{\rm s}$ &
  $-68\arcdeg 59\arcmin 30\arcsec$ & 0.54 & 1.4 & 0.3 & $-0.92$ \\
197 & $5^{\rm h} 39^{\rm m} 50^{\rm s}$ &
  $-70\arcdeg 07\arcmin 00\arcsec$ & 0.57 & 4.7 & 0.05 & $-1.50$ \\
216 & $5^{\rm h} 44^{\rm m} 35^{\rm s}$ &
  $-69\arcdeg 26\arcmin 00\arcsec$ & 0.47 & 4.9 & 0.2 & $-0.83$ \\
225 & $5^{\rm h} 47^{\rm m} 20^{\rm s}$ &
  $-70\arcdeg 41\arcmin 30\arcsec$ & 0.67 & 1.1 & 0.1 & $-1.55$ \\
\enddata
\tablenotetext{a}{At
$3\arcsec \times 3\arcsec \times 1\;{\rm km\,s^{-1}}$ resolution}
\tablenotetext{b}{A CO-to-${\rm H_2}$ conversion factor of
$7\times 10^{20}\;{\rm cm^{-2}\,(K\,km\,s^{-1})^{-1}}$
is assumed \citep{fukui2008}}
\tablenotetext{c}{The fraction of the CO luminosity in the ALMA FoV
over that of the entire GMC \citep{fukui2008}}
\end{deluxetable*}

\begin{figure*}
\plotone{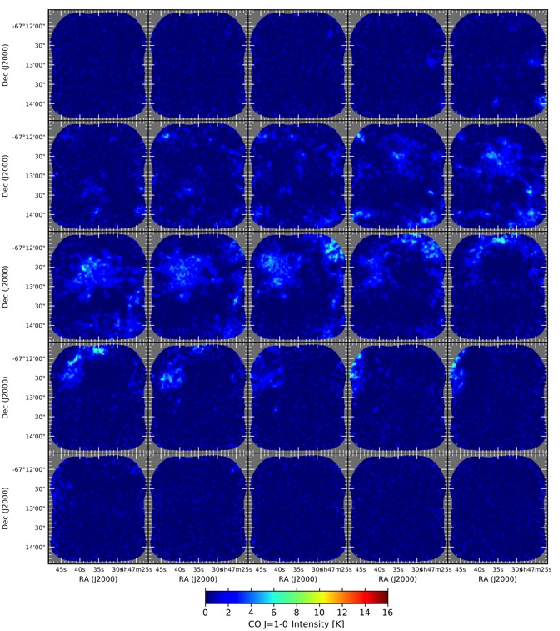}
\caption{Velocity channel maps of GMC 2 at $v_{\rm LSR} = 246$,
247, ..., $270\;{\rm km\,s^{-1}}$.}
\label{fig:chmap002}
\end{figure*}

\begin{figure*}
\plotone{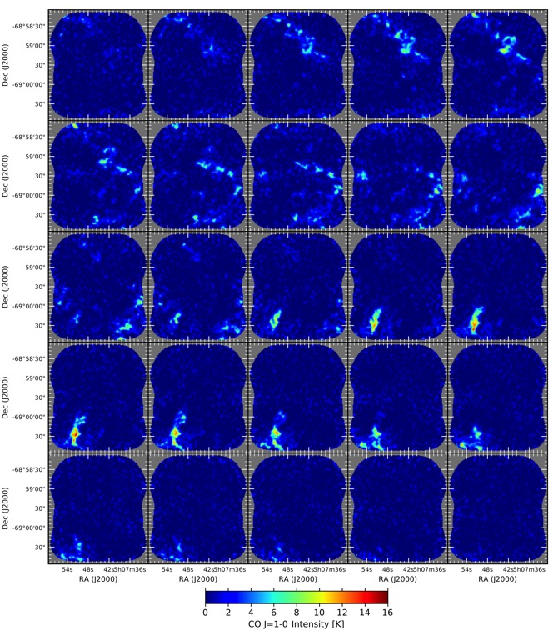}
\caption{Velocity channel maps of GMC 55 at $v_{\rm LSR} = 246$,
247, ..., $270\;{\rm km\,s^{-1}}$.}
\label{fig:chmap055}
\end{figure*}

\begin{figure*}
\plotone{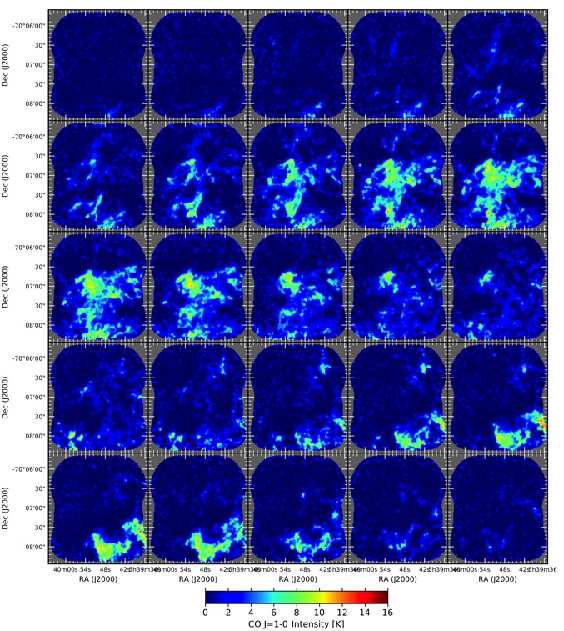}
\caption{Velocity channel maps of GMC 197 at $v_{\rm LSR} = 221$,
222, ..., $245\;{\rm km\,s^{-1}}$.}
\label{fig:chmap197}
\end{figure*}

\begin{figure*}
\plotone{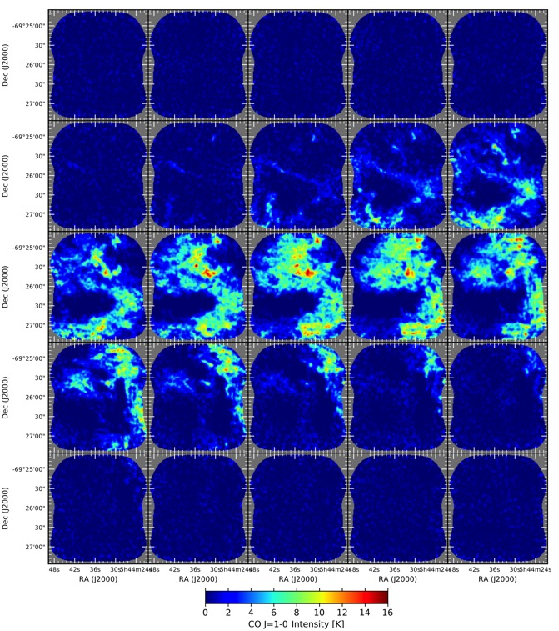}
\caption{Velocity channel maps of GMC 216 at $v_{\rm LSR} = 216$,
217, ..., $240\;{\rm km\,s^{-1}}$.}
\label{fig:chmap216}
\end{figure*}

\begin{figure*}
\plotone{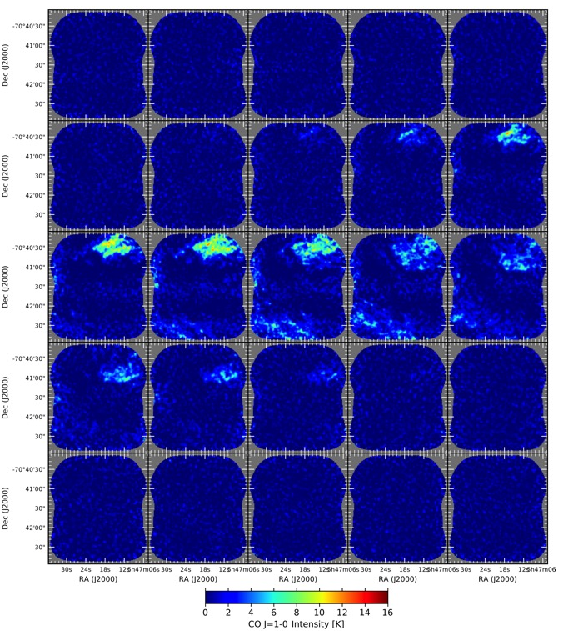}
\caption{Velocity channel maps of GMC 225 at $v_{\rm LSR} = 206$,
207, ..., $230\;{\rm km\,s^{-1}}$.}
\label{fig:chmap225}
\end{figure*}

\begin{figure*}
\plotone{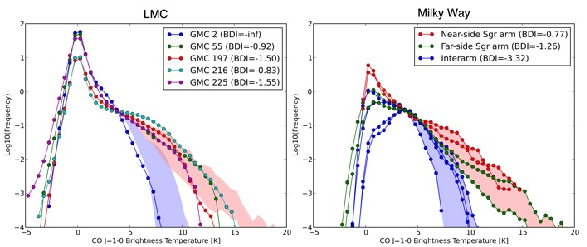}
\caption{Left: the brightness distribution functions (BDFs) of GMCs 2, 55,
197, 216, and 225.
The frequency is normalized by the number of the pixels with 3--5 K
intensity.
The red and blue shaded regions are the BDFs of the
spiral arm and inter-arm regions of the Milky Way (see the right panel).
Right: the BDFs of the gas in the Milky Way.
The data \citep{sawada2012} are smoothed to match the spatial
resolution of the ALMA data (0.73 pc).
Four 5 ${\rm km\,s^{-1}}$ velocity bins were taken for the arm
($v_{\rm LSR}=40\mbox{--}45$, $45\mbox{--}50\;{\rm km\,s^{-1}}$
for the nearside Sgr arm; 55--60, $60\mbox{--}65\;{\rm km\,s^{-1}}$
for the farside Sgr arm) and inter-arm regions
(75--80, 80--85, 85--90, and $90\mbox{--}95\;{\rm km\,s^{-1}}$),
respectively.}
\label{fig:BDF}
\end{figure*}

\section{Discussion} \label{sec:discussion}

We observed five GMCs in the LMC along the GMC evolutionary stages
classified by \citet{kawamura2009}.
As discussed in Section \ref{sec:targets}, additional information
became available after this work started,
showing YSOs in GMC 55 and 225, which were originally classified
as Type I (no sign of massive star formation).
Among the five, only GMC 2 does not show any sign
of star formation. In addition, it turned out that our FoVs do
not optimally cover the target GMCs. 
Even though our ALMA mosaics are large
($2\farcm 6 \times 2\farcm 8$; or $38\times 41\;{\rm pc^2}$),
the emission peaks at the $45\arcsec$ resolution \citep{wong2011}
were considerably offset from
the peaks at the $2\farcm 6$ resolution \citep{fukui2008}.
With these limitations in mind, there still seems to be a trend of
GMC evolution, when our observations are viewed in the context of
work in the literature.

\subsection{The Brightness Distribution Function and Index} 
\label{sec:BDF}

The maps of the five GMCs (Figure \ref{fig:intandmax1}) clearly show
less developed internal structures in GMC 2 than in the others.
GMC 2 is the one without any sign of star formation, 
while the others have some associated YSOs.

In order to characterize the visual difference in internal structures,
\citet{sawada2012} introduced simple tools:
the BDF and BDI.
The BDF is a histogram of the brightness of a line emission,
which quantitatively represents the map appearance.
It is defined in $\alpha$--$\delta$--$v$ space as the fraction
of the ``pixels'' with brightness between $T$ and $T+dT$.
The BDI is a single number that represents the characteristics of the
BDF: the fraction of the bright, structured gas over
the bulk unstructured, faint gas.
It is written as the flux ratio of the bright emission to faint
emission:
\begin{eqnarray}
{\rm BDI} &=& \log_{10} \left(
  \frac{\int_{T_2}^{T_3} T\cdot B(T) dT}
       {\int_{T_0}^{T_1} T\cdot B(T) dT}
 \right) \nonumber\\
 &=& \log_{10} \left(
  \frac{\sum_{T_2<T[i]<T_3}T[i]}{\sum_{T_0<T[i]<T_1}T[i]}
 \right),
 \label{eq:bdi}
\end{eqnarray}
where $B(T)$ denotes the BDF;
$T_0$, $T_1$, $T_2$, and $T_3$ are the brightness thresholds.
$T[i]$ is the brightness of the $i$th pixel in
the $\alpha$--$\delta$--$v$ space.
In this paper, we adopt 
$(T_0, T_1, T_2, T_3) = (3, 5, 10, \infty)$ [K],
the same brightness thresholds used for the MW analysis
in \citet{sawada2012}.
A high BDI indicates that the gas is structured and
has more compact, bright structures.

The BDF/BDI analysis revealed that in the MW disk,
structured gas, represented by the high BDI
($\sim -1$ in CO $J=1\mbox{--}0$ seen at $\simeq 0.7$ pc resolution),
is distributed along the spiral arms, while unstructured
(low BDI; $\sim -3$) gas exists in the inter-arm regions.
Thus, there is an evolution
in gas structure across the spiral arms.
Figure \ref{fig:BDF} (right) shows the BDFs of the molecular
gas in the MW around the Galactic longitude of
$38\arcdeg$ from \citet{sawada2012}.
Assuming the locations of the Sagittarius arm (near and far sides)
and inter-arm regions as discussed in \citet{sawada2012},
the spatial scales are smoothed to match the 0.7 pc resolution
of this LMC study. The BDFs of the spiral arms show a tail toward
high brightness temperature, while the inter-arm molecular
gas does not show the tail.
The red and blue shadings are displayed for comparisons with
the GMCs in the LMC below.
The BDF/BDI analysis has been also applied to external galaxies
\citep{hughes2013}.

We applied the BDF/BDI analysis to the five GMCs.
The data are smoothed to a $2\;{\rm km\,s^{-1}}$
velocity resolution to achieve better sensitivity.
The resultant spatial (0.7 pc) and velocity resolutions and sensitivity
(0.37--0.48 K) are comparable to the MW study \citep{sawada2012}.
Figure \ref{fig:BDF} (left) shows the BDF of the GMCs.
The red and blue shadings are the same as those in the right panel;
the red indicates the locus of the BDFs of the spiral arms in the MW,
while the blue is for the inter-arm region.
GMC 2 is the only GMC without any sign of star formation,
and its structure is similar to the quiescent molecular gas
in the MW inter-arm regions.
Its BDI is very low ($-\infty$; meaning no gas with high brightness
$>10$ K).
The other four GMCs show some signs
of star formation, and their BDFs, $-1.5$ to $-0.8$,
are similar to those of the spiral arms in the MW.
The BDIs of the GMCs are shown in the figure legend and
Table \ref{tbl:results}.
Although the correlation exists between the high BDI and star
formation activity, the cause of the complex structure in
molecular clouds is debatable as to whether they are a precursor
of star formation or developed by stellar feedback.
We, however, note that in the MW high BDI regions do not
necessarily coincide with \ion{H}{2} regions, and hence
indicating the former \citep[see the next subsection]{sawada2012let}.

\subsection{Synthesis with Work in the Literature}

Using the BDF/BDI, \citet{sawada2012let} demonstrated
that in the MW, structured gas, represented by a high BDI,
is distributed along the spiral arms, while unstructured
(low BDI) gas exists in the inter-arm regions
(see Figure \ref{fig:BDF} right).
The overall distribution of the high BDI gas is similar to that
of the \ion{H}{2} regions.
They also found some moderately high BDI regions,
which are massive molecular concentrations and located in the
spiral arms, but without \ion{H}{2} regions.
These regions may be in the phase between quiescent and active
star-forming GMCs.
Compact, and presumably dense, structures
have already developed in molecular gas, but in which
star formation has not yet started.
They may be pre-star-forming complexes.

Similar structural differences are present among local star-forming
and quiescent molecular clouds.
\citet{kainulainen2009} analyzed the probability distribution
function (PDF) of column density derived from the near-infrared
dust extinction.
They found that the quiescent clouds show a log-normal shape of the PDF
as predicted for a turbulent medium,
while the star-forming clouds show excess ``wings'' at higher
column densities, presumably due to the self-gravity of dense regions
\citep{klessen2000,federrath2008}.
The PDF and BDF appear very similar if the CO brightness is converted to
column density using the CO-to-${\rm H_2}$ conversion factor.
In fact, the PDFs of star-forming and quiescent clouds
are {\it quantitatively} consistent with the BDFs of arm and inter-arm
molecular gas, respectively \citep{sawada2012, sawada2012let}.
A caveat is that it is not guaranteed that the conversion factor is
applicable on a pixel-by-pixel basis.
Other optically thin lines, e.g., $^{13}{\rm CO}$ or ${\rm C^{18}O}$,
may trace dense gas better than $^{12}{\rm CO}$, but may miss extended,
diffuse gas components (i.e., the denominator of the BDI).
It is also possible that the emission lines may not necessarily trace
the column/volume density of the gas, but rather the excitation of the
spectral lines.
We, however, note that \citet{sawada2012} found a consistent result
with $^{12}{\rm CO}$ and $^{13}{\rm CO}$ in the BDF analysis in the MW.

Most ALMA studies have been focused on GMCs with active star formation
(Figure \ref{fig:overall}).
A quantitative analysis of their data is beyond the scope of
this paper, but qualitatively we find a similar trend:
structured gas in star-forming clouds and less structured in
quiescent clouds.
\citet{indebetouw2013} observed the northern part of 30 Doradus,
the most active star-forming region in the Local Group.
In this region, a large fraction of the molecular gas -- about 50\% --
has a brightness temperature above 10 K (see their Figure 4).
This cloud has already developed dense structures.
In our five GMCs, only the densest clumps have such high brightness
temperatures.
N159 is another active star-forming region near 30 Doradus
that has also been observed
with ALMA \citep{fukui2015, saigo2017, nayak2018}.
The presence of dense clumps and filaments led the authors
to speculate on collisions of clouds or filaments as a possible
cause of the structures.
N55 is an \ion{H}{2} region in the largest SGS in the LMC.
A cloud near N55 is also very clumpy \citep{naslim2018}.
The same is true in the star-forming cloud N83C in the SMC
\citep{muraoka2017}.

\citet{wong2017} presented the first ALMA study to contrast
star-forming and quiescent molecular clouds in the LMC.
They observed a quiescent cloud, designated as the Planck Galactic Cold
Clump (PGCC) G282.98$-$32.40 \citep[][Wong et al.\ called this the
``Planck cold cloud'' or PCC]{planck2016}.
It does not show any signs of active massive star formation,
and the estimated dust temperature is low ($T_{\rm d}\sim$15 K).
The CO brightness distribution shows a much lower contrast than
that in the 30 Doradus cloud and is relatively uniform 
with only one hotspot.
The quiescent cloud is less structured.

By synthesizing our five GMCs, one quiescent and four star-forming,
with those in the literature,
a simple view of a GMC evolutionary sequence is tempting:
clouds start from a relatively structureless state,
develop complex internal structures, and, as a result,
form stars.

\subsection{Caveats}

Of course, this evolutionary sequence in star formation activity
may be too simplistic, as other factors, such as environmental triggers,
might be at a play.
For example, interactions with SGSs may be another important factor
\citep{meaburn1980,kim1999,yamaguchi2001,book2009,dawson2013,fujii2014}.
In fact, GMCs 55, 197, and 216 show relatively high BDIs and 
are located at the edges of SGS complexes identified by
\citet{dawson2013}.
The question here is what is the direct cause of the gas with
high brightness temperature (perhaps, dense gas).
If compression by an SGS directly changes the cloud structure
(hence, BDF or PDF) without invoking gas self-gravity, we may need to
consider its effects separately from the internal physics.
The other possible factor is surrounding GMCs.
Figure \ref{fig:intandmax1} (left) shows that
some of the GMCs in \citet{fukui2008} are GMC complexes rather
than single GMCs. These are some of the reasons why even the relatively
large ALMA mosaics ($\approx40$ pc) are not enough to cover
the entire emission. If the surrounding GMCs trigger compression,
e.g., by tides or collisions, the environment would be another factor
to consider.
Studies into multiple factors clearly require a much
larger sample of GMCs in the LMC. Such statistics should
be possible with ALMA as it is approaching its operational maturity.

\acknowledgments
This paper makes use of the following ALMA data:
ADS/JAO.ALMA\#2012.1.00641.S.
ALMA is a partnership of ESO (representing
its member states), NSF (USA) and NINS (Japan), together with NRC
(Canada), MOST and ASIAA (Taiwan), and KASI (Republic of Korea),
in cooperation with the Republic of Chile.
The Joint ALMA Observatory is operated by ESO, AUI/NRAO and NAOJ.
This research made use of Astropy, a community-developed core 
Python package for Astronomy \citep{astropy2013}.
This research made use of Montage.
It is funded by the National Science Foundation under
grant number ACI-1440620, and was previously funded by
the National Aeronautics and Space Administration's
Earth Science Technology Office,
Computation Technologies Project,
under Cooperative Agreement Number NCC5-626
between NASA and the California Institute of Technology.
J.K.\ acknowledge the support from the NSF under grant AST-1211680.
We thank J.\ Barrett for improving the manuscript.

\vspace{5mm}
\facilities{ALMA, Mopra, {\it Spitzer}}

\software{CASA \citep{mcmullin2007}, Miriad \citep{sault1995}, 
Astropy \citep{astropy2013}}

\listofchanges

\end{document}